\documentstyle[12pt]{article}
	 \makeatletter
	 \def\imi{|I_iM_i>}
	 \def\imf{|I_fM_f>}
	 \def\ifmf{<I_fM_f|}
	 \def\phirt{\phi({\bf r}', \ t)}
	 \def\e0{\hat{\bf e}_0}
	 \def\emu{\hat{\bf e}_\mu}
	 \def\bgq{\begin{equation}}
	 \def\edq{\end{equation}}
	 \def\bga{\begin{eqnarray}}
	 \def\eda{\end{eqnarray}}
	 \def\sc{\scriptstyle }
	 \def\thefigure{\@arabic\c@figure}\def\fps@figure{tbp}
	 \def\ftype@figure{1}\def\ext@figure{lof}
	 \def\fnum@figure{\protect\footnotesize Fig.\ \thefigure}
	 \def\thetable{\@arabic\c@table}
	 \def\fps@table{tbp}\def\ftype@table{2}\def\ext@table{lot}
	 \def\fnum@table{\protect\footnotesize Table \thetable}
	 \makeatother
\begin{document}
\vspace*{0.3in}
\begin{center}
{\Large {\bf
Theory of Multiphonon Excitation in Heavy-Ion Collisions
}}\\[1cm]

{\Large
  C.A. Bertulani, L.F. Canto} \\
  \bigskip
  Instituto de F\'\i sica, Universidade Federal do Rio de Janeiro\\
  Cx. Postal 68528\\
  21945-970 Rio de Janeiro, RJ, Brazil\\
  \bigskip
  \bigskip
  {\Large M.S. Hussein$^+$}  \\
\bigskip

		    Center for Theoretical Physics\\
		    Laboratory for Nuclear Science and Department of Physics \\
		 Massachusetts Institute of Technology, \
	       Cambridge, Massachusetts, 02139,  USA $^*$\\
				and\\
	  Institute for Theoretical Atomic and Molecular Physics\\
	   at the Harvard-Smithsonian Center for Astrophysics \\
	   60 Graden St. Cambridge, Massachusetts, 02138, USA $^\ddagger$\\

  \bigskip\bigskip
{\Large A.F.R. de Toledo Piza}\\
  \bigskip
  Instituto de F\'\i sica, Universidade de S\~ao Paulo\\
  Cx. Postal 66318, 05389-970 S\~ao Paulo, SP, Brazil\\

\bigskip
\end{center}
\vfil
\noindent + Permanent address: Instituto de F\'\i sica,
Universidade de S\~ao Paulo, Cx. Postal 66318, 05389-970 S\~ao
Paulo, SP, Brazil. Supported in part by FAPESP.
\medskip

\noindent * Supported
in part by funds provided by the U.S. Department of Energy
(D.O.E.) under cooperative agreement
DE-FC02-94ER40818
\medskip

\noindent $\ddagger$ Supported by the National Science
Foundation
\newpage

\centerline{\bf ABSTRACT}
\begin{quotation}
\vspace{-0.10in}

We study the effects of channel coupling in the excitation dynamics of
giant resonances in relativistic heavy ions collisions. For this purpose,
we use a semiclassical approximation to the Coupled-Channels problem and
separate the Coulomb and the nuclear parts of the coupling into their
main multipole components. In order to assess the importance of multi-step
processes, we neglect the resonance widths and solve the set of coupled
equations exactly. Finite widths are then considered. In this case, we
handle the coupling of the ground state with the dominant Giant Dipole
Resonance exactly and study the excitation of the remaining resonances
within the Coupled-Channels Born Approximation. A comparison with recent
experimental data is made.
\end{quotation}
\vskip 1cm

\baselineskip 4ex

\section{Introduction}

Relativistic Coulomb Excitation (RCE) is a well established tool
to unravel interesting aspects of nuclear structure \cite{baur}.
Examples are the studies of multiphonon resonances
in the SIS accelerator at the GSI facility, in Darmstadt,
Germany \cite{hans,chomaz}. Important properties of
nuclei far from stability \cite{PG} have also been studied with this method.
The RCE induced by large-Z projectiles and/or targets, often yields
very large cross sections, which
depend on the nuclear response to the
acting electromagnetic fields.  In such cases, one expects
a strong coupling between the excited states. This coupling might
be responsible for the large discrepancies between experimental
data of RCE and the calculations based on first-order
perturbation theory~\cite{baur,hans,chomaz}, or the harmonic
oscillator model.

In the present paper, we apply a semiclassical method~\cite{AW65} to the
coupled-channels (CC) problem and study RCE in several collisions between
heavy ions. In this method, the projectile-target relative
motion is approximated by a classical trajectory and the
excitation of the Giant Resonances is treated quantum mechanically~\cite{Ca94,
Be11}. The use of this method is justified due to the small wavelenghts
associated with the relative motion. In section 2, we neglect the resonance
widths and introduce the semiclassical CC-equations for relativistic Coulomb
excitation. The
time--dependent matrix-elements of the main multipole components of the
Coulomb (section 2.1) and nuclear (section 2.2) parts of the coupling
interaction are calculated. The CC-equations are then solved (section 2.3)
in some limiting cases. Section 3 is devoted to the excitation of resonances
of finite widths. Generalizing the schematic treatment of ref.~\cite{Ca94},
we present an ``exact" solution for the coupling between the g.s. and the
dominant GDR.  The excitation of the weaker resonances are then  evaluated
through the Coupled-Channels Born Approximation (CCBA), from the g.s. and
GDR amplitudes. In section 4 we apply the results of the previous sections to
specific cases and make a comparison with recent  experimental data. Finally,
in section 5, we summarize our results and present the conclusions of this
work.

\bigskip\bigskip

\section{The semiclassical method for the CC-problem}

In relativistic heavy ion collisions, the wavelength associated to the
projectile-target separation is much smaller than the
characteristic lengths of system. It is, therefore, a reasonable
approximation to treat ${\bf r}$ as a classical variable ${\bf
r}(t)$, given at each instant by the trajectory followed by the
relative motion. At high energies, is also a good approximation
to replace this trajectory by a straight line.  The intrinsic
dynamics can then be handled as a quantum mechanics problem with
a time dependent hamiltonian. This treatment is discussed in full
details by Alder and Winther in ref.~\cite{AW65}.

The intrinsic state $\vert \psi(t)>$ satisfies the Schr\"odinger equation
\begin{equation}
\left[ h\ +\ V({\bf r}{\scriptstyle (t)}) \right]\ \vert \psi(t)> =
i \hbar {\partial \vert \psi(t)>
\over \partial t}\; . \label{eqS}
\end{equation}
\noindent Above, $h$ is the intrinsic hamiltonian and $V$ is the
channel-coupling interaction.

Expanding the wave function in the set $\{ \vert m>;\ m=0,N\}$ of eigenstates
of $h$, where $N$ is the number of excited states included in the
Coupled-Channel problem, we obtain
\begin{equation}
\vert \psi (t)> = \sum_{m=0}^N\ a_m(t)\ \vert m>\ \exp \Big(-i
E_m t /\hbar\Big) \; , \label{expan}
\end{equation}
\noindent
where $E_m$ is the energy of the state $\vert m>$. Taking scalar product
with each of the states $<n \vert $, we get the set of coupled equations
\begin{equation}
i \hbar\ {\dot a}_n(t) = \sum_{m=0}^N\ < n\vert V\vert m >
\;
e^{i(E_n-E_m) t /\hbar}\; a_m(t)
\qquad \qquad n =0\; {\rm to}\;\;  N \,.\label{AW}
\end{equation}

\noindent It should be remarked that the amplitudes depend also
on the impact parameter $b$ specifying the classical trajectory followed
by the system. For the sake of keeping the notation simple, we do not
indicate this dependence explicitly. We write, therefore, $a_n(t)$ instead
of $a_n(b,t)$. Since the interaction $V$ vanishes as $t\rightarrow \pm
\infty$, the amplitudes have as initial condition $a_n(t\rightarrow -\infty)
= \delta(n,0)$ and they tend to constant values as
$t\rightarrow \infty$. Therefore, the excitation probabity of an intrinsic
state $\vert n> $ in a collision with impact parameter $b$ is given as
\begin{equation}
P_n(b) = \vert a_n(\infty)\vert ^2\; .\label{Pn}
\end{equation}

\noindent The total cross section for excitation of the state $\vert n>$ can
be approximated by the classical expression
\begin{equation}
\sigma_n=2\pi\ \int\ P_n(b)\ b db \;. \label{sigman}
\end{equation}

Since we are interested in the excitation of
specific nuclear states, with good angular momentum and parity quantum
numbers, it is appropriate to develop the time-dependent coupling interaction
$V(t)$ into multipoles. In ref.~\cite{WA79}, a multipole expansion of the
electromagnetic excitation amplitudes in relativistic heavy ion collisions
was carried out. This work used first order perturbation theory and the
semiclassical approximation. The time-dependence of the multipole
interactions was not explicitly given. In section 2.1 we show how this
time-dependence can be explicitly obtained, from the
Taylor-series expansion of the Li\'enard-Wiechert potentials \cite{Ja75} and
the continuity equation for the nuclear current.

In section 2.2 we deduce the time-dependence and the
multipole decomposition of the nuclear interaction in relativistic
nucleus-nucleus collisions. The nuclear absorption at collisions below
grazing impact parameter is also accounted for.

\subsection{Coulomb excitation}
\medskip

We consider a nucleus 1 which is at rest and a relativistic
nucleus 2 which moves along the $z$-axis and is excited from the
initial state $\imi$  to the state $\imf$ by the electromagnetic
field of nucleus 1.  The nuclear states are specified by the
spin quantum numbers $I_i$, $I_f$ and by the corresponding
magnetic quantum numbers $M_i$ and $M_f$, respectively. We
assume that the relativistic nucleus 2 moves along a
straight-line trajectory with impact parameter $b$, which is
therefore also the distance of the closest approach between the
center of mass of the two nuclei at the time $t=0$. We shall
consider the situation where $b$ is larger than the sum of the
two nuclear radii, such that the charge  distributions of the
two nuclei do not strongly overlap at any time.  The
electromagnetic field of the nucleus 2 in the reference frame of
nucleus 1 is given by the usual Lorentz transformation
\cite{Ja75} of the scalar potential $\phi({\bf r})= Z_1e/|{\bf
r}|$, i.e.,
\begin{eqnarray}
\phi({\bf r}', t) &=& \gamma \ \phi\Big[{\bf b'-b}, \ \gamma (z'-vt)\Big],
\nonumber \\
{\bf A}({\bf r}', t) &=&
{{\bf v}\over c}\ \gamma \ \phi\Big[{\bf b'-b}, \ \gamma (z'-vt)\Big].
\label{Aphi}
\end{eqnarray}
Here ${\bf b}$ (impact parameter) and ${\bf b}'$ are the components
of the radius-vectors ${\bf r}$  and
${\bf r}'$ transverse to ${\bf v}$.

The time-dependent matrix element for electromagnetic excitation
is of the form
\begin{equation}
V_{fi}(t)=\ifmf \Big[ \rho({\bf r}')-{{\bf v}\over c^2} \cdot {\bf J}
({\bf r}') \Big] \ \phi({\bf r}', \ t) \imi \ .\label{Vfi}
\end{equation}

A Taylor-series expansion of the Li\'enard-Wiechert
potential around ${\bf r}'=0$ yields
\begin{equation}
\phirt=\gamma \phi[{\bf r}(t)]+\gamma \nabla \phi[{\bf r}(t)]\cdot {\bf r}'
+ \cdots\label{phiexp}
\end{equation}
where  ${\bf r}=({\bf b}, \ \gamma v t)$, and the following
simplifying notation is used:
\begin{eqnarray}
\nabla \phi[{\bf r}]& \equiv& \nabla' \phirt
\bigg|_{{\bf r}'=0}
\nonumber\\
&=& - \nabla_{\bf b} \phi({\bf r}) - {\partial \over \partial (vt)}
\phi({\bf r}) \ \hat{\bf z} =
- \nabla_{\bf b} \phi({\bf r}) - {{\bf v} \over  c^2} \
{\partial \over \partial t} \phi({\bf r})\ .\label{nphi}
\end{eqnarray}
Thus,
\begin{equation}
V_{fi}(t)=\ifmf \Big[ \rho({\bf r}')-{{\bf v}\over c^2} \cdot {\bf J}
({\bf r}') \Big] \
\Big[ \gamma \phi({\bf r})+\gamma {\bf r}'\cdot \nabla \phi({\bf r})
\Big]\imi \ .\label{Vfi1}
\end{equation}
Using the continuity equation
\bgq
\nabla \cdot {\bf J}=-i \ \omega \ \rho\ ,\label{ce}
\edq
where $\omega=(E_f-E_i)/\hbar$, and integrating by parts,
\begin{equation}
V_{fi}(t)=\ifmf \Big[ {\bf J}({\bf r}) \cdot \Big[
{\nabla' \over i\omega} -{{\bf v} \over c^2}\Big]
\
\Big[ \gamma \phi({\bf r}+\gamma {\bf r}'\cdot \nabla \phi({\bf r})
\Big]\imi \ .\label{Vfi2}
\end{equation}

In spherical coordinates
\bgq
{\bf r}' \cdot \nabla \phi={\sqrt{4\pi}\over 3} \
\sum_{\mu=-1}^1 \alpha_\mu \ r'\ Y_{1\mu}^*
\ ,\label{rgrad}
\edq
where
\bgq
\alpha_\mu=\hat{e}_\mu \cdot \nabla \phi\label{alpha}
\edq
We will use the relations
\bgq
{{\bf r}\over c^2}={v\over c^2} \hat{e}_0 =
{v\over c^2} \ \sqrt{4\pi \over 3} \
\nabla (rY_{10}^*) \label{rc2}
\edq
and
\bgq
\nabla \times {\bf L} (r^k Y_{lm})=i(k+1) \nabla
(r^kY_{lm})\label{nL}
\edq
where ${\bf L} = -i \nabla \times {\bf r}$.

Then, one can write
\begin{eqnarray}
{\bf J}\cdot \Big({\nabla \over i\omega} -{{\bf v}\over c^2}\Big)
\ \Big[ \gamma \phi +\gamma {\bf r}' \cdot \nabla \phi\Big]&=&
- \gamma {\bf J} \cdot \bigg[
{{\bf v}\over c^2} (\nabla \phi \cdot {\bf r}')
\ -\ \sqrt{4\pi \over 3} \nonumber\\
&\times& \Big\{
\sum_{\mu=-1}^1 {\alpha_\mu\over i\omega} \
\nabla' (r'Y_{1\mu})-{v\over c^2} \
\phi \ \nabla' (r'Y_{10}^*)\Big\}\
\bigg]
\ .\label{J}
\end{eqnarray}
The last term in the above equation can be rewritten as
\bgq
\Big({\bf J}\cdot {{\bf v} \over c^2}
\Big) \ \Big({\bf r}'\cdot \nabla \phi\Big)
= {v\over 2c^2} \ {\bf J} \cdot
\Big[ \e0 ({\bf r}'\cdot \nabla \phi)+({\bf r}'\cdot \e0)\nabla \phi \Big]
%\nonumber \\
+ {v\over 2c^2} \ {\bf J} .
\Big[ \e0 ({\bf r}'\cdot \nabla \phi)-({\bf r}'\cdot \e0)\nabla \phi
\Big]
\ .\label{Jvc}
\edq
The first term in this equation is symmetric under parity inversion,
and contributes to the electric quadrupole (E2) excitation amplitudes,
since
\bgq
 {v\over 2c^2} \ {\bf J} \cdot
\Big[ \e0 ({\bf r}'\cdot \nabla \phi)+({\bf r}'\cdot \e0)\nabla \phi
\Big]=
 {v\over 2c^2} \ {\bf J} \cdot \nabla' \Big[ z' ({\bf r}' \cdot \nabla \phi)
 \Big]
 \ .\label{v2c2}
 \edq
 The second term in eq.~(\ref{Jvc}) is antisymmetric in $\bf J$ and $\bf r'$,
 and leads to magnetic dipole (M1) excitations. Indeed, using
 eqs.~(\ref{rgrad}--\ref{nL}), one finds
 \bgq
 {v\over 2c^2} \ {\bf J} \cdot
\Big[ \e0 ({\bf r}'\cdot \nabla \phi)-({\bf r}'\cdot \e0)\nabla \phi
\Big]=
 {v\over 2c^2} \ {\bf J} \cdot  \Bigg[
\sqrt{4\pi \over 3} \sum_{\mu=-1}^1 \alpha_\mu (-1)^\mu \
{\bf L} \Big({\bf r} Y_{1,-\mu}\Big)
 \Bigg]
 \ .\label{v2c3}
 \edq
Thus, only the first two terms on the right-hand-side
of eq.~(\ref{J}) contribute to the
electric dipole (E1) excitations. Inserting them into eq.~(\ref{Vfi2}), we get
\bgq
V^{(E1)}_{fi}(t)=\gamma \ \sqrt{4\pi\over 3} \
\sum_{\mu=-1}^1
(-1)^\mu \ \beta_\mu \
\ifmf {\cal M} (E1, -\mu) \imi
\ ,\label{Vfi3}
\edq
where
\bgq
{\cal M} (E1, -\mu) = {i\over \omega} \ \int d^3r {\bf J}({\bf r})
. \nabla \Big(rY_{1\mu}\Big)= \int d^3 r \ \rho({\bf r}) \ r \ Y_{1\mu}
({\bf r})
\ ,\label{ME1}
\edq
and
\bga
\beta_{\pm}&=&-\alpha_\mu = - \big( \nabla \phi
\cdot \emu \big) = \emu \cdot {\partial \phi \over \partial {\bf b}} \nonumber
\\
\beta_0 &=& -\alpha_0 - i {\omega v \over c^2} \phi
\ .\label{beta}
\eda
The derivatives of the potential $\phi$ are explicitly
given by
\bga
{\partial \phi \over \partial {\bf b}_x} &\equiv&
\nabla_{{\bf b}_x} \phi \Big|_{{\bf r}'=0}= - \hat{\bf x}  \
b_x {Z_1e \over [b^2+\gamma^2v^2t^2]^{3/2}}\nonumber \\
\nabla_z \phi\Big|_{{\bf r}'=0}&=& - \hat{\bf z} \ \gamma^2 v t \
{Z_1e\over [b^2+\gamma^2v^2t^2]^{3/2}}
\ .\label{deriv}
\eda

Using the results above, we get for the electric dipole
potential
\bga
V^{(E1)}_{fi}(t)&=& \sqrt{2\pi\over 3}\ \gamma \
\Bigg\{ {\cal E}_1 (\tau) \ \Big[ {\cal M}_{fi} (E1, -1) -
{\cal M}_{fi} (E1, 1)\Big]
\nonumber \\
&+&\sqrt{2} \ \gamma \ \tau\ \bigg[ {\cal E}_1(\tau)
-
i{\omega v b\over c^2}
\ \Big( 1+\tau^2 \Big)\  {\cal E}_2(\tau)
\bigg] \ {\cal M}_{fi} (E1, 0) \Bigg\}
\ ,\label{Vfi4}
\eda
where $\tau=\gamma v/b$, and
\bgq
{\cal E}_1(\tau)={ Z_1 e  \over
b^2 \ [1+\tau^2]^{3/2}} \
\ \ \ \
{\rm and} \  \ \
{\cal E}_2(\tau)={ Z_1 e \tau  \over b \ [1+\tau^2]^{3/2}} \label{E1}
\edq
are the transverse and longitudinal eletric fields
generated the relativistic nucleus with charge $Z_1e$, respectively.
{}From the definition
\bgq
{\cal M}_{fi} (M1, \mu)= - {i\over 2c}\ \int d^3r \ {\bf J}
({\bf r}) . {\bf L} \Big( r Y_{1\mu}\Big)
\ ,\label{Mfi}
\edq
and eq.~(\ref{v2c2}),  we find
\bgq
V^{(M1)}_{fi}(t)= i \ \sqrt{2\pi\over 3}\ {v\over c} \
{\cal E}_1 (\tau) \ \Big[ {\cal M}_{fi} (M1, 1)
-
{\cal M}_{fi} (M1, -1)\Big]
\ .\label{Vfi5}
\edq

To obtain the electric quadrupole (E2) potential we use the
third term in the Taylor expansion of eq.~(\ref{phiexp}). Using the
continuity equation, a part of this term will contribute to E3
and M2 excitations, which we neglect. We then find that
\bga
V^{(E2)}_{fi}(\tau)&=& -\sqrt{\pi\over 30}\ \gamma \
\bigg\{ 3 \ {\cal E}_3 (\tau) \
\Big[ {\cal M}_{fi} (E2, 2) +
{\cal M}_{fi} (E2, - 2)\Big] \nonumber \\
&+&\gamma \Big[ 6\ \tau  {\cal E}_3(\tau) - i \
{\omega v\over \gamma c^2} \Big(1+\tau^2\Big)
\ {\cal E}_1(\tau) \Big]
\ \Big[ {\cal M}_{fi} (E2, -1) +
{\cal M}_{fi} (E2, 1)\Big]\nonumber \\
&+&
\sqrt{6} \ \gamma^2 \ \Big[ \Big( 2 \tau^2-1\Big)
\ {\cal E}_3(\tau) - i \ {\omega v \over \gamma c^2} \
\tau \ \Big(1+\tau^2\Big)\ {\cal E}_1(\tau)\Big]
{\cal M}_{fi} (E2, 0)
\bigg\}
\ ,\label{Vfi6}
\eda
where
${\cal E}_3(\tau)$ is
the quadrupole electric field of nucleus 1, given by
\bgq
{\cal E}_3(\tau)={ Z_1 e
\over b^3\ [1+\tau^2]^{5/2}}
\ .\label{E3}
\edq

The fields ${\cal E}_i(\tau)$ peak around $\tau=0$, and
decrease fastly within an interval
$\Delta \tau \simeq 1$.
This corresponds to a collisional
time $\Delta t \simeq b / \gamma v$.
This means that, numerically one needs to integrate
the Coupled-Channels equations (eq.~(\ref{AW})) only in a
time interval within a range $n\times \Delta \tau$ around
$\tau=0$, with
$n$ equal to a small integer number. This will be shown latter in
connection with the calculation presented in section 4.

Using the Wigner-Eckart theorem we can write \cite{Edmonds}
\bgq
{\cal M}_{fi}(E\lambda, \mu) =
(-1)^{I_f-M_f} \
\biggl( {I_f \atop -M_f}{\lambda \atop \mu}
{I_i \atop M_i} \biggr)
\ifmf | {\cal M} (E\lambda) |\imi \ .\label{Mfi1}
\edq

A  phase convention for the nuclear states
can be found so that the reduced matrix elements
$\ifmf | {\cal M} (E\lambda) |\imi$ are
real numbers \cite{AW65}. For the case of
giant resonances, sum rules are very useful to guess
the values of these matrix elements. It is usual to
use the reduced transition probability
\bga
B(E\lambda; I_i \longrightarrow I_f )
&=& {1\over 2I_i+1}\
\sum_{M_i M_f} \Big| <I_iM_i| {\cal M} (E\lambda, \ \mu)
|I_fM_f>\Big|^2 \nonumber \\
&=&
{1\over 2I_i+1}\ \Big| <I_i|| {\cal M} (E\lambda)
||I_f>\Big|^2 \ ,\label{BEl}
\eda
in terms of which the energy-weighted sum-rules
yield, for the E1 and E2 excitations,
\begin{equation}
B(E1;\ I_i \longrightarrow I_f)=
\bigg({1\over 2I_f+1}\bigg)\
{9 \over 4\pi} \ {\hbar^2 \over 2 m_N} \
{NZ\over AE_x} \ e^2
\ ,\label{BEif}
\end{equation}
and
\begin{equation}
B(E2; \
I_i \longrightarrow I_f)=
\bigg( {1\over 2I_f+1}\bigg)\
{\hbar^2\over m_N} \ {15R^2\over 4\pi E_x} \ e^2
\times \left\{ \begin{array}{ll}
    Z^2/A, & \mbox{ for isoscalar excitations; }\\
   NZ/A, & \mbox{for isovector excitations; }
\end{array}
\right. \label{BE2}
\end{equation}
where $N$, $Z$, and $A$ are the neutron, charge, and mass
number of the excited nucleus, respectively.
In these equations it was assumed that an isolated state
with energy $E_x$ exhausts the sum-rule.

The matrix elements for the transitions between multiphonon
states can be determined by
using the Wigner-Eckart theorem and the reduced matrix
elements inferred from sum rules, as described in sections
2.1 and 2.2. In the case of perfect phonons, i.e.,
eigenstate solutions of the harmonic oscillator, the
following relation holds for the reduced
matrix elements for the transition
$0 \rightarrow 1$ and $n-1 \rightarrow n$ \cite{BZ93}:
\bgq
\vert <n-1||V_{E/N,1}||n>\vert^2=n\ \vert <0||V_{E/N,1}||1>\vert^2 \ .
\edq
The factor $n$ on the r.h.s. is the boson enhacement factor.

\subsubsection{Approximate solutions}

In most cases, the first-order perturbation theory is a good
approximation to calculate the amplitudes for relativistic
Coulomb excitation. It amounts to using $a_k=\delta_{k0}$ on the right
hand side of
eq.~(\ref{AW}). The time integrals can be evaluated analytically
for the $V_{Ei}(t)$ perturbations, given by eqs. (\ref{Vfi4}),
(\ref{Vfi5}), and (\ref{Vfi6}).  The result is
\bga
a^{(E1)}_{1st}&=& - i \sqrt{8\pi\over 3}\ {Z_1 e \over
\hbar v b} \ \xi \ \
\Bigg\{ K_1(\xi) \ \Big[ {\cal M}_{fi} (E1, -1) -
{\cal M}_{fi} (E1, 1)\Big]
\nonumber \\
&+& i {\sqrt{2} \over \gamma} \ K_0(\xi) \
{\cal M}_{fi} (E1, 0) \Bigg\}
\ ,\label{a1st}
\eda
where $K_1$ ($K_2$) is the modified Bessel function of first (second) degree,
and $\xi=\omega b/\gamma v$. For the E2 and M1 multipolarities, we
obtain respectively,
\bga
a^{(E2)}_{1st}&=& 2 i \ \sqrt{\pi\over 30}\ {Z_1 e \over
\gamma \hbar vb^2} \ \xi^2
\bigg\{ K_2(\xi)\
\Big[ {\cal M}_{fi} (E2, 2) +
{\cal M}_{fi} (E2, - 2)\Big] \nonumber \\
&+&i \gamma \ \Big(2-{v^2\over c^2}\Big) \ K_1(\xi)
\ \Big[ {\cal M}_{fi} (E2, -1) +
{\cal M}_{fi} (E2, 1)\Big]\nonumber \\
&-&
\sqrt{6} \
K_0(\xi) \
{\cal M}_{fi} (E2, 0)
\bigg\}
\ ,\label{aE2}
\eda
and
\bgq
a^{(M1)}_{1st}= \sqrt{8\pi\over 3}\ {Z_1e\over \hbar c b} \ \xi
\ K_1(\xi) \ \Big[ {\cal M}_{fi} (M1, 1)
-
{\cal M}_{fi} (M1, -1)\Big]
\ .\label{aM1}
\edq

These expressions are the same as those obtained from the
formulae deduced in ref. \cite{WA79}. We note that the multipole
decomposition developed by those authors is accomplished by a
different approach, i.e., using recurrence relations for the
Gegenbauer polynomials, after the integral on time is performed.
Therefore, the above results present a good check for the
time-dependence of the multipole fields deduced here.

A simplified model, often used in connection with multiphonon excitations,
is the harmonic vibrator model. In this model, the resonance widths are
neglected and the Coupled-Channel equations
can be solved exactly, in terms of the first-order excitation amplitudes
\cite{baur}. The excitation amplitude of the $n$-th harmonic oscillator
state, for any time $t$, is given by
\bgq
a_{h.o.}^{(n)} (t) ={\Big[ a_{1st} (t) \Big]^n \over \sqrt{n!}} \
\exp \Big\{ -|a_{1st} (t)|^2/2 \Big\}\ ,\label{aho}
\edq
where $a_{1st} (t) $ is the excitation amplitude for the
$0 \ (g.s.)\longrightarrow 1\ (one \ phonon)$
calculated with the first-order perturbation
theory.

For the excitation of giant resonances, $n$ can be identified
with the state corresponding to a multiple $n$ of the single
giant resonance state.  This procedure has been often used in
order to calculate the cross sections for the excitation of
multiphonon giant resonances.  Since this result is exact in the
harmonic vibrator model, it accounts for all coupling between
the states. However, this result can be applied to studies of
giant resonance excitation only if the same class of multipole
states is involved.  I.e., if one considers only electric dipole
excitations, and use the harmonic oscillator model, one can
calculate the excitation probabilities, and cross sections, of
the GDR, double-GDR, triple-GDR, etc.  Eq. (\ref{aho}) is not
valid if the excitation of other multipolarities are involved,
e.g., if the excitation of dipole states  and quadrupole states
are treated simultaneously.  In ref. \cite{Norbury} a hybrid
harmonic oscillator model has been used. In this work, it is assumed
that the difference between the amplitudes obtained with the
harmonic oscillator model and with $n$-th order perturbation theory
is due to the appearence of the exponential term on
the r.h.s. of eq.~(\ref{aho}). This exponential takes care of
the decrease in the  occupation amplitude of the ground state as
a function of time. As argued in ref. \cite{Norbury}, the
presence of other multipole states, e.g., of quadrupole states,
together with dipole states, may be accounted for by adding the first
order excitation amplitudes for the quadrupole states to the
exponent in eq.~(\ref{aho}). This would correct for the flux
from the ground state to the quadrupole states. In other words,
eq.~(\ref{aho}) should be corrected to read
\bgq
a_{h.o.}^{(n)} (\pi\lambda,\ t) ={\Big[ a_{1st} (\pi\lambda, \ t)
\Big]^n \over \sqrt{n!}} \
\exp \Big\{ - \sum_{\pi'\lambda'} \Big|
a_{1st} (\pi'\lambda', \ t)\Big|^2/2 \Big\}\ .\label{ahon}
\edq

The harmonic oscillator model is not in complete agreement with
the experimental findings. The double-GDR and double-GQR
states do not have exactly twice the energy of the respective
GDR and GQR states \cite{hans,chomaz}. Apparently, the matrix
elements for the transition from the GDR (GQR) to the double-GDR
(double-GQR)
state does not follow the boson-rule \cite{BZ93} (see end of section 3).
This is borne out by the discrepancy between the experimental cross
sections for the excitation of the double-GDR and the double-GQR with the
perturbation theory, and with the harmonic oscillator model
\cite{hans,chomaz}. Thus, a Coupled-Channels calculation is useful
to determine which matrix elements for the transitions among the
giant resonance states reproduce the experimental data.

\subsection{Nuclear excitation and strong absorption}
\medskip
In peripheral collisions the nuclear interaction between the ions can
also induce excitations. This can be easily calculated in a vibrational
model.
The amplitude for the
excitation of a vibrational mode
by the nuclear interaction in
relativistic heavy ion collisions
can be obtained assuming that
a residual interaction $U$ between the projectile and the target exists,
and that it is weak. According to the Bohr-Mottelson
particle-vibrator
coupling model, the matrix element for the transition $i\longrightarrow
f$ is given by
\begin{equation}
V^{N(\lambda\mu)}_{fi} ({\bf r})
\equiv \ifmf U \imi=
{\delta_\lambda \over \sqrt{2\lambda +1}}
\ \ifmf Y_{\lambda \mu}\imi
\ Y_{\lambda \mu} (\hat {\bf r})
\ U_{\lambda} (r) \label{VfiN}
\end{equation}
where $\delta_\lambda=\beta_\lambda R$
is the vibrational amplitude, or {\it
deformation length},
$R$ is the nuclear radius, and $U_{\lambda}
(r)$ is the transition potential.

The deformation length $\delta_\lambda$ can be directly related to the
reduced matrix elements for electromagnetic transitions. Using well-known
sum-rules for these matrix elements one finds a relation between the
deformation length
and the nuclear masses and sizes.
For isoscalar excitations one gets \cite{Sa87}
\begin{equation}
\delta_0^2= 2 \pi \ {\hbar^2 \over m_N\ <r^2>} \
{1 \over A E_x} \ , \ \ \ \
\delta_{\lambda \geq 2}^2 = {2 \pi \over 3} \ {\hbar^2 \over m_N} \
\lambda \ (2\lambda +1) \ {1\over A E_x}\label{d0}
\end{equation}
where $A$ is the atomic
number, $<r^2>$ is the r.m.s. radius of the nucleus,
and $E_x$ is the excitation energy.

The transition potentials for isoscalar exciations are
\begin{equation}
U_0 (r) = 3 U_{opt} (r) + r {d U_{opt} (r) \over dr}
\ ,\label{U0}
\end{equation}
for monopole, and
\begin{equation}
U_2 (r)= {dU_{opt} (r) \over dr} \ , \label{U2}
\end{equation}
for quadrupole modes.

For dipole isovector excitations
\begin{equation}
\delta_1= {\pi \over 2} \ {\hbar^2 \over m_N}
\ {A \over NZ} \ {1\over E_x}\ , \label{d1}
\end{equation}
where $Z$ ($N$) the charge (neutron)  number. The transition potential
in this case is \cite{Sa87}
\begin{equation}
U_1(r)=\chi \ \Big( {N-Z \over A} \Big) \
\Big( {dU_{opt} \over dr} + {1\over 3} \ R_0 \ {d^2 U_{opt}
\over dr^2} \Big)
\ ,\label{U1}
\end{equation}
where the factor
$\chi$ depends on the difference between the proton and the neutron
matter radii  as
\begin{equation}
\chi {2(N-Z)\over 3A} = {R_n-R_p \over {1\over 2} \ (R_n+R_p)}
= {\Delta R_{np} \over R_0}
\ .\label{chi}
\end{equation}
Thus, the strength of isovector excitations increases with
the difference
between the neutron and the proton matter radii.
This difference is accentuated for neutron-rich
nuclei and should be a good test for the quantity $\Delta R_{np}$
which enters the above equations.

The time dependence of the matrix elements above can be obtained
by making a Lorentz boost, assuming that $U_\lambda$ is the scalar
part of a four-vector with zero vector-potential.
Since the potentials
$U_\lambda\Big[r(t)\Big]$ peak strongly
at $t=0$, we can safely approximate $\theta (t)
\simeq \theta(t=0) =\pi/2$ in the
spherical harmonic of eq.~(\ref{VfiN}). One gets
\bga
V^{N(\lambda\mu)}_{fi} ({\bf r})
&\equiv& \ifmf U \imi
\nonumber \\
&=&\gamma \
{\delta_\lambda \over \sqrt{2\lambda +1}}
\ \ifmf Y_{\lambda \mu}\imi
 Y_{\lambda \mu} \Big(\theta={\pi\over 2}\Big)
\ U_{\lambda} [r(t)]  \ ,\label{VfiN2}
\eda
where
$
r(t)=\sqrt{b^2+\gamma^2 v^2 t^2}
$.

Using the Wigner-Eckart theorem, the matrix element of the
spherical harmonics becomes
\bgq
\ifmf Y_{\lambda \mu}\imi =
(-1)^{I_f-M_f} \
\biggl[ {(2I_i+1)(2\lambda+1) \over 4 \pi (2I_f+1)}
\biggr]^{1/2} \
\biggl( {I_f \atop -M_f}{\lambda \atop \mu}
{I_i \atop M_i} \biggr)
\biggl( {I_f \atop 0}{\lambda \atop 0}
{I_i \atop 0} \biggr)
\ .\label{WE}
\edq

For high energy collisions, the optical potential $U(r)$
can be constructed by using the t-$\rho\rho$ approximation
\cite{HRB91}. One gets
\bgq
U(r)=-{\hbar v \over 2} \ \sigma_{NN} \ (\alpha_{NN}+i)
\ \int \rho_1({\bf r}')\ \rho_2({\bf r-r'})\ d^3 r'
\ ,\label{Ur}
\edq
where $\sigma_{NN}$ is the nucleon-nucleon cross section, and
$\alpha_{NN}$ is the real-to-imaginary ratio of the forward
($\theta=0^\circ$) nucleon-nucleon scattering amplitude.
A set of the experimental values of these quantities,
useful for our purposes, is given in table 1.

We are not interested here in diffraction and
refraction effects in the scattering, but on the excitation
probabilities for a given impact parameter.
The strong absorption occuring in collisions with small
impact parameters can be included. This can be done
by using the eikonal approximation and the
optical potential, given by eq.~(\ref{Ur}).
The practical result is that the excitation probabilities
for a given impact parameter $b$, including the sum of the
nuclear and the Coulomb contributions to the excitation,
are given by
\bgq
P_{fi}(b)= \Big| a_{fi}^C(b)+
a_{fi}^N(b) \Big|^2 \ \exp\Big\{
- \sigma_{NN}
\ \int dz \int d^3r \rho_1({\bf r}')\
\rho_2({\bf r-r'})
\Big\}
\ ,\label{abs}
\edq
where
$r=\sqrt{b^2+z^2}$. The corresponding excitation
cross sections are obtained by an integration of the
above equation over impact parameters.

\bigskip

\section{The effect of finite resonance widths}

Up to now we have assumed that the excited states are isolated
states, with zero width. However, this assumption is not
realistic and it is important to study the effect of finite
resonance widths on the excitation amplitudes.  This is
specially relevant for the case of excitation of giant
resonances, which have a broad structure. The simplest way to
study this effect is by using the Coupled-Channels Born
approximation. This approximation was used ref. \cite{Ca94} to
describe the excitation of the double giant resonance in
relativistc heavy ion collisions.  It is based on the idea that
in such cases only the coupling between the ground state and the
dominant giant dipole state has to be treated  exactly.  The
reason is that the transitions to giant quadrupole and to the
double-phonon states have low probability amplitudes, even for
small impact parameters.  However, an exact treatment of the
back-and-forth transitions between the ground state and the
giant dipole state is necessary. This leads to
modifications of the transitions amplitudes to the remaining
resonances, which are populated by the ground state and the GDR.
In ref. \cite{Ca94} the application of the method
was limited to the use of an schematic interaction, and the
magnetic substates were neglected.  These deficiencies are
corrected here. The method allows the inclusion of the width of
the giant resonances in a very simple and straightforward way.
It will be useful for us to compare with the Coupled-Channels
calculations with isolated states, as we described in the
previous sections.  Figure~1 represents our procedure. The GDR
is coupled to the ground state while the remaining resonaces are
fed by these two states according to first order perturbation
theory. The coupling matrix elements involves the ground state and
a set of doorway states $|D^{\sc (n)}_{\sc \lambda\mu}>$, where $n$
specifies the kind of resonance and $\lambda\mu$ are angular momentum
quantum numbers. The amplitudes of these resonances in real continuum
states are
\bgq
\alpha^{\sc (n)}(\epsilon)=<\phi(\epsilon)\Big|D^{\sc (n)}_{\sc \lambda\mu}>
\label{aeps}
\ ,
\edq
where $\phi(\epsilon)$ denotes the wavefunction of one
of the numerous states which are responsible for the broad
structure of the resonance.  In this equation
$\epsilon=E_x-E_n$, where $E_x$ is the excitation energy and
$E_n$ is the centroid of the resonace considered.

As we have stated above, in this approach we use the
Coupled-Channels equations for the coupling between the
ground state and the GDR. This results in the following
Coupled-Channels equations:

\bga
i\hbar \ {\dot a}_{\sc 0}(t) &=&
\sum_\mu \ \int d\epsilon <\phi(\epsilon) |D^{\sc (1)}_{\sc 1\mu}>
\ <D^{\sc (1)}_{\sc 1\mu}|V_{E1,\mu}(t) |0> \ \exp\Big\{ -
{i\over \hbar} (E_{\sc 1}+\epsilon)t\Big\}
\ a^{\sc (1)}_{\sc \epsilon,1\mu}(t)\nonumber \\
&=&
\sum_\mu \ \int d\epsilon \ \alpha^{\sc (1)}(\epsilon)
\ V^{\sc (01)}_{\sc \mu}(t) \ \exp\Big\{ -
{i\over \hbar} (E_{\sc 1}+\epsilon)t \Big\}
\ a^{\sc (1)}_{\sc \epsilon,1\mu}(t)\ ,\label{da0}
\eda
and
\bgq
i\hbar \ {\dot a}^{\sc (1)}_{\sc \epsilon,1\mu}(t) =
\Big[ (\alpha^{\sc (1)}(\epsilon)\ V^{\sc (01)}_{\sc \mu}(t)\Big]^* \
\exp\Big\{ i (E_1+\epsilon)t/\hbar \Big\}
\ a_{\sc 0}(t)\ .\label{da1}
\edq
Above, $(n=1)$ stands for the GDR, $a_{\sc 0}$ denotes the occupation
amplitude of the ground state and $a^{\sc (1)}_{\sc \epsilon,1\mu}$
the occupation amplitude of a state located at an
energy $\epsilon$ away from the GDR centroid, and with
magnetic quantum number $\mu$ ($\mu = -1,0,1$).
We used the short hand notation $V^{\sc (01)}_{\sc \mu}(t)=<D^{\sc
(1)}_{\sc 1\mu}|V_{E1,\mu}(t) |0>$.

Integrating eq.~(\ref{da1}) and inserting the result in
eq.~(\ref{da0}), we get the integro-differential equation for the
ground state occupation amplitude
\begin{eqnarray}
{\dot a}_{\sc 0}(t)&=& - \frac{1}{\hbar^2}\ \sum_\mu \
V^{\sc (01)}_{\sc \mu}(t)\
\int
d\epsilon\  |\alpha^{\sc (1)}(\epsilon)|^2 \nonumber\\
& \times &
\int_{-\infty}^t dt'\
\left[ V^{\sc (01)}_{\sc \mu}(t')\right]^* \ \exp
\Big\{ -i (E_{\sc 1}+\epsilon)
(t-t')/\hbar\Big\} \ a_{\sc 0}(t')\ ,\label{da01}
\end{eqnarray}
where we used that $a^{\sc (1)}_{\sc \epsilon,1\mu}(t=-\infty)=0$.
To carry out the integration over $\epsilon$, we should use
an appropriate parametrization for the doorway amplitude
$\alpha^{\sc (1)}(\epsilon)$. A convenient choice is the
Breit-Wigner (BW) form
\bgq
|\alpha^{\sc (1)}(\epsilon)|^2 = { 1 \over 2 \pi}\
\left[ {\Gamma_{\sc 1} \over \epsilon^2+\Gamma_{\sc 1}^2/4 }\right]
\ .\label{aeps2}
\edq
In this case, this integral will be the simple exponential
\bgq
\int d\epsilon \ |\alpha^{\sc (1)}(\epsilon)|^2 \
\exp\Big\{
-i {(E_1+\epsilon)t\over \hbar}\Big\}
=\exp\Big\{ -i{(E_1-i\Gamma/2) t\over\hbar}\Big\}
\ .
\edq

A better agreement with the experimental line shapes of the
giant resonances is obtained by using a Lorentzian (L)
parametrization for $|\alpha^{\sc (1)}(\epsilon)|^2$, i.e.,
\bgq
|\alpha^{\sc (1)}(\epsilon)|^2 = {2 \over \pi}
\ \left[ {\Gamma_{\sc 1}\ E_x^2 \over (E_x^2-E_{\sc 1}^2)^2+
\Gamma_{\sc 1}^2 E_x^2 }\right]
\ ,\label{L}
\edq
where $E_x=E_{\sc 1}+\epsilon$.
The energy integral can still be performed exactly~\cite{Pato} but now
it leads to the more complicated result
\bgq
\int d\epsilon \ |\alpha^{\sc (1)}(\epsilon)|^2\ \exp\Big\{
-i {(E_{\sc 1}+\epsilon)t\over \hbar} \Big\}
=\Big(1-i{\Gamma_{\sc 1}\over 2E_{\sc 1}} \Big) \
\exp\Big\{ -i{(E_{\sc 1}-i\Gamma_{\sc 1}/2) t\over\hbar}\Big\}\
+\ \Delta C(t)\ ,
\edq
where $\Delta C(t)$ is a non-exponential correction to the decay.
For the energies and widths involved in the excitation of giant
resonances, this correction can be shown numerically to be negligible.
It will therefore be ignored in our subsequent calculations.
After integration over $\epsilon$, eq.~(\ref{da01}) reduces to

\bgq
{\dot a}_{\sc 0}(t)\ =-\ {\cal S}_1\
\sum_\mu \ V^{\sc (01)}_{\sc \mu} (t)  \int_{-\infty}^t dt'\
\left[ V^{\sc (01)}_{\sc \mu}(t')\right]^*  \exp \Big\{ -i {(E_{\sc
1}-i\Gamma_{\sc 1}/2)
(t-t') \over \hbar}\Big\} \ a_{\sc 0}(t')\ ,\label{da02}
\edq
where the factor ${\cal S}_1$ is ${\cal S}_1 = 1$ for BW-shape and
${\cal S}_1 =  1 - i \Gamma_{\sc 1}/2 E_{\sc 1}$ for L-shape.

We can take advantage of the exponential time-dependence in the integrand
of the above equation, to reduce it to a set of second order differential
equations. Introducing the auxiliary amplitudes $A_{\sc \mu}(t)$, given
by the relation
\bgq
a_{\sc 0}(t)= 1\ +\ \sum_\mu A_{\sc \mu} (t) \ ,\label{A}
\edq
with initial conditions $A_{\sc \mu} (t=-\infty) = 0$, and
taking the derivative of eq.~(\ref{da02}), we get
\bgq
{\ddot A}_{\sc \mu}(t) - \bigg[ {{\dot V}^{\sc (01)}_{\sc \mu} (t) \over
V^{\sc (01)}_{\sc \mu} (t)}\  -\
{i\over \hbar} \Big( E_{\sc 1} -i {\Gamma_{\sc 1} \over 2}
\Big)\bigg] \ {\dot A}_{\sc \mu}(t)\
-\ {\cal S}_1\ {|V^{\sc (01)}_{\sc \mu}(t)|^2 \over \hbar^2} \ \Big[
1+\sum_{\mu'}A_{\mu'}(t)\Big] \ .
\edq

Solving the above equation, we get $a_{\sc 0}(t)$. Using this amplitude
and integrating eq.~(\ref{da1}), one can evaluate $a^{\sc (1)}_{\sc
\epsilon,1\mu}(t)$. The probability density for the population of a
GDR continuum state with energy $E_x$ in a collision with impact parameter $b$,
$P_1(b,E_x)$, is obtained trough the summation over
the asymptotic ($t\rightarrow \infty$) contribution from each magnetic
substate. We get

\bgq
P_1(b,\ E_x) = |\alpha^{\sc (1)}(E_x-E_{\sc 1})|^2
\ \sum_\mu \bigg| \int_{-\infty}^\infty dt'\
\exp\Big\{iE_xt'\Big\} \ \left[ V^{\sc (01)}_{\sc \mu}(t')\right]^* \ a_0(t')
\bigg|^2\ ,\label{pro1}
\edq
where $|\alpha^{\sc (1)}(E_x-E_{\sc 1})|^2$ is given by eq.~(\ref{aeps2}) or
by eq.~(\ref{L}), depending on the choice of the resonance shape.

To first order, DGDR continuum states can be populated through $E2$-coupling
from the ground state or through $E1$-coupling from GDR states.
The probability density arising from the former is given by eq.~(\ref{pro1}),
with the replacement of the line shape $|\alpha^{\sc(1)}|^2$ by its DGDR
counterpart $|\alpha^{\sc(2)}|^2$
(defined in terms of parameters $E_{\sc 2}$ and $\Gamma_{\sc 2}$)
and the use of the appropriate coupling-matrix elements $V^{\sc
(02)}_{\sc \mu}(t)$ with the E2 time dependence given by (29).
On the other hand, the contribution from the latter
process is
\bga
P_2(b,E_x)& =& |\alpha^{\sc (2)}(E_x-E_{\sc 2})|^2\, {\cal S}_1\
\ \sum_\nu \bigg| \int_{-\infty}^\infty dt' \
\exp\Big\{iE_xt'\Big\} \ \Big\{\sum_\mu \ \left(V^{\sc (12)}_{\sc \nu\mu}
(t')\right)^*
\nonumber \\
&\times& \int_{-\infty}^{t'} dt''\
\ \left( V^{\sc (01)}_{\sc \mu}(t'')\right) \
\exp \Big\{ -i {(E_{\sc 1}- i \Gamma_{\sc 1}/2)
(t-t')\over \hbar} \Big\} \ a_{\sc 0}(t'')
\bigg|^2 \ , \label{P2}
\eda
We should point out that eq.~(\ref{P2}) is {\bf not} equivalent to
second-order perturbation theory. This would be  true only in the limit
$a_0(t) \longrightarrow 1$. In our approach, $a_0(t)\ne 1$, since it
is modified by the time-dependent coupling to the GDR state.
This coupling is treated exactly by means of the
Coupled-Channels equations. We consider that this is the main
effect on the calculation of the DGDR excitation probability.
This approach is justified due to the small
excitation amplitude for the transition $1 \longrightarrow 2$,
since $a_1(t) \ll a_0(t)$.

Equations  similar to (\ref{pro1}) can also be used to calculate the
ISGQR and IVGQR excitation probabilities, with the proper choice
of energies, widths, and transition potentials (e.g., $V_{E2}(t)$, or
$V_{N2}(t)$, or both).

In the next section we will apply the results of sections
2.1, 2.2 and 3, to analyse some examples of relativistic
nuclear and Coulomb excitation.

\bigskip

\section{Applications}

We consider the excitation of giant resonances in $^{208}$Pb
projectiles, incident on $^{208}$Pb targets at 640 A$\cdot$MeV.
This reaction has been recently studied at the GSI/SIS,
Darmstadt~\cite{hans}. For this system the excitation
probabilities of the isovector giant dipole ($IVGD$) at 13.5 MeV
are large and, consequently, high order effects of channel coupling
should be relevant. To assess the importance of these
effects, we assume that the GDR state  depletes 100\% of
the energy-weighted sum-rule and neglect the resonance width.
The influence of resonance widths will be considered later, in section 4.2.

\subsection{Zero-width calculations}

As a first step, we study the time evolution of the excitation
process, solving the Coupled-Channels equations for a reduced
set of states.  We consider only the ground state (g.s.) and the
GDR. The excitation probability is then compared with that
obtained with first order perturbation theory. This is done in
figure~2, where we plot the occupation probabilities of the
g.s., $|a_0(t)|^2$, and of the GDR, $|a_1(t)|^2$, as functions
of time, for a collision with impact parameter $b= 15$ fm. As
discussed earlier, the Coulomb interaction is strongly peaked
around $t=0$, with a width of the order $\Delta t \simeq
b/\gamma v$. Accordingly, the amplitudes are rapidly varying in
this time range. A comparison between the CC-calculation (solid
line) and first order perturbation theory (dashed line) shows
that the the high order processes contained in the former lead
to an appreciable reduction of the GDR excitation probability.
{}From this figure we can also conclude that our numerical
calculations can be restricted to the interval $-10 < \tau <10,$
where $\tau = (\gamma v/b)\ t$ is the time variable measured in
natural units. Outside this range, the amplitudes reach
assymptotic values.

It is worthwhile to compare the predictions of first order
perturbation theory with those of the harmonic oscillator model
and the CC calculations.  In addition to the GDR, we include the
following multiphonon states: a double giant dipole state
($2\otimes IVGD$) at 27 MeV, a triple giant dipole state
($3\otimes IVGD$) at 40.5 MeV, and a quadruple giant dipole
state ($4\otimes IVGD$) at 54 MeV. The coupling between the
multiphonon states are determined by  boson factors, as
explained at the end of section~2.2. Direct excitations of the
multiphonon states from the g.s.  are not considered. The
angular momentum addition rules for bosons yields the following
angular momentum states: $L=0$ and $2$, for the $2\otimes GDR$
state; $L=1$, 2, and 3, for the $3\otimes GDR$ state; and $L=0$,
1, 2, 3, and 4, for the $4\otimes GDR$ state. We assume that
states with the same number of phonons are degenerate. In table
2, we show the resulting cross sections. The excitation
probabilities and the cross section were calculated with the
formalism of section~2. The integration over impact parameter
was carried out in the interval $b_{\sc min} < b < \infty$.  As we
discuss below, the low-$b$ cut-off value~\cite{BZ93} $b_{\sc
min} = 14.3$~fm mocks up absorption effects. We have checked that
the CC results are not significantly affected by the unknown
phases of the transition matrix elements. Since the multiphonon
spectrum is equally spaced, and the coupling matrix-elements are
related through boson factors (see the end of section~2.2), the
harmonic oscillator and the CC cross sections should be equal.
In fact the numerical results of these calculations given in the
table are very close.  We also see that the excitation cross
sections of triple- and quadruple-phonon states are much smaller
than that for the $2\otimes GDR$.  Therefore, we shall
concentrate our studies on the $2\otimes GDR$, neglecting other
multiphonon states.

Next, we include the remaining important giant resonances in
$^{208}$Pb.  Namely, the isoscalar giant quadrupole ($ISGQ$) at
10.9~MeV and the isovector giant quadrupole ($IVGQ$) at 22~MeV.
Also in this case, we use 100 \% of the energy-weighted sum
rules to deduce the strength matrix elements.  In table~3, we
show the excitation probabilities in a grazing collision, with
$b=14.3$~fm. We see that first order perturbation theory yields
a very large excitation probability for the $IVGD$ state. This
is strongly reduced in a c.c.  calculation, as we have already
disscussed in connection with the figure 2. The excitations of
the remaining states are also influenced. They are reduced due
to the lowering of the occupation probabilities of the g.s. and
of the $IVGD$ state in the c.c.  calculation. As expected,
perturbation theory  and  c.c. calculations agree at large
impact parameters, when the transition probabilities are small.
For the excitation of the $2\otimes IVGD$ state we used
second-order perturbation theory to obtain the value in the
second column. The presence ot the ISGQR and the IVGQR influence
the c.c. probabilities for the excitation of the GDR and the
$2\otimes IVGD$, respectively.

We should also consider the effects of strong absorption in grazing
collisions, as discussed in section~2.2.
In figure~3 we plot the GDR excitation probability as a function
of the impact parameter. In the solid line, we consider absorption
according to eq.~(\ref{abs}). In the construction of the optical
potential we used the g.s. densities calculated from the droplet model of
Myers and Swiatecki~\cite{MS}. As shown in ref.~\cite{baur}, this
parametrization yields the best agreement between experiment and theory.
The dashed line does not include absorption. To simulate strong absorption
at low impact parameters, we use $b_{\sc min} = 15.1$~fm as a lower limit
in the impact parameter integration of eq.~(\ref{sigman}). This value was
chosen such as to lead to the same cross section as that obtained from the
solid line.

In figure 4, we plot the nuclear contributions to the excitation
probability, and as a function of the impact parameter. We study
the excitation of the isoscalar giant monopole resonance
(ISGMR), the IVGDR, and the ISGQR. The ISGMR in $^{208}$Pb is
located at 13.8 MeV.  As discussed previously, isovector
excitations are suppressed in nuclear excitation processes, due
to the approximate charge independence of the nuclear
interaction.  We use the formalism of section 2.2, with the
deformation parameters such that  100\% of the sum rule is
exhausted. This corresponds to the monopole amplitude
$\alpha_0=0.054$. The IVGDR and ISGQR deformation parameters are
$\delta_1=0.31$ fm and $\delta_2=0.625$ fm, respectively.  The
IVGQR excitation probability is much smaller than the other
excitation probabilities and is, therefore, not shown. The
nuclear excitation is peaked at the grazing impact parameter and
is only relevant within an impact parameter range of $\sim 2$
fm. Comparing to figure~3, we see that these excitation
probabilities are orders of magnitude smaller than those for
Coulomb excitation. Consequently, the corresponding cross
sections are much smaller. We get 14.8 mb for the isoscalar GDR,
2.3 mb for the ISGQR, and 2.3 mb for the IVGDR. The interference
between the nuclear and the Coulomb excitation is also small and
can be neglected.

\subsection{Effect of resonance widths}

We now turn to the influence of the giant resonance widths on the
excitation dynamics. We use the CCBA
formalism developed in section 3. Schematically, the CC problem is that
represented figure~1. As we have seen above, the strongest coupling occurs
between the g.s. and the GDR.

In figure~5, we show the excitation energy spectrum for the GDR, the DGDR
(a short hand notation for the $2\otimes$IVGD), ISGQR and IVGQR. The centroid
energies and the widths of these resonances are listed in Table~4. The
figure show excitation spectra obtained with both Breit-Wigner (BW) and
Lorentzian (L) line shapes. One observes that the BW and L spectra have similar
strengths at the resonance maxima. However, the low energy parts (one or two
widths below the centroid) of the spectra are more than one order of magnitude
higher in the BW calculation. The reason for this behavior is that Coulomb
excitation favors low energy transitions and the BW has a larger low energy
tail as compared with the Lorentzian line shape. The contribution from the
DGDR leads to a pronounced bump in the total energy spectrum. This bump depends
on the relative strength of the DGDR with respect to the GDR. In figure~6, we
show the ratio $\sigma_{\sc DGDR}/\sigma_{\sc GDR}$ as a function of the
bombarding energy. We observe that this ratio is roughly constant in the
energy range $E_{\sc lab}/A = 200\ -\ 1000$~MeV and it falls beyond these
limits. This range corresponds to the SIS-energies at the GSI-Darmstadt
facility.

We now study the influence of the resonance widths and shapes on the GDR
and DGDR cross sections. This study is similar to that presented in
ref.~\cite{Ca94}, except that we now have a realistic three dimensional
treatment of the states and consider different line shapes. In the upper part
figure~7, denoted by (a), we show $\sigma_{\sc GDR}$ as a function of
$\Gamma_{\sc GDR}$, treated as a free parameter. We note that the BW and L
parametrizations lead to different trends. In the BW case the cross section
grows with $\Gamma_{\sc GDR}$ while in the L case it decreases. The growing
trend is also found in ref.~\cite{Ca94}, which uses the BW line shape. The
reason for this trend in the BW case is that an increase in the GDR width
enhances the low energy tail of the line shape, picking up more
contributions from the low energy transitions, favored in Coulomb excitation.
On the other hand, an increase of the GDR width enhances the doorway amplitude
to higher energies where Coulomb excitation is weaker. In figure~7 (b) and (c),
we study the dependence of $\sigma_{\sc GDR}$ on $\Gamma_{\sc GDR}$. In (b),
the DGDR width is kept fixed at the value 5.7~MeV while in (c) it is kept
proportional to $\sigma_{\sc GDR}$, fixing the ratio $\Gamma_{\sc DGDR} /
\Gamma_{\sc GDR} = \sqrt{2}$. The first point to be noticed is that
the BW results are sistematically higher than the L ones. This is a
consequence of the different low energy tails of these functions, as
discussed above. One notices also that $\sigma_{\sc DGDR}$ decreases
with $\Gamma_{\sc GDR}$ both in the BW and L cases. This trend can be
understood in terms of the uncertainty principle. If the GDR width is
increased, its life-time is reduced. Since the DGDR is dominantly populated
from the GDR, its short life-time leads to decay before the transition to the
DGDR.

To assess the sensitivity of the DGDR cross section on the strength of the
matrix elements and on the energy position of the resonance, we
present in table 5 the cross sections for the excitation of the GDR,
DGDR, ISGQR and IVGQR, obtained with the CCBA approximation and 100\% of
the sum-rules for the respective modes.
In this calculation we have included
the strong absorption, as explained in section 2.2.
For comparison, the values inside
parenthesis (and brackets) of the DGDR excitation cross section include a
direct excitation of the L=2 DGDR state. We assumed that 20\% of the
E2 sum rule could be allocated for this excitation mode of the DGDR. The
cross sections increase by less than 10\% in this case.
The value inside parenthesis (brackets) assume a positive (negative)
sign of the matrix element for the direct excitation.

Since the excitation of the DGDR is weak, it is very well described
by eq.~(\ref{P2}) and the DGDR population is approximately
proportional to the squared strength of $V^{\sc (12)}$.
Therefore, to increase the DGDR cross section by a factor of 2,
it is necessary violate the relation of eq.~(35) by the same
factor.  This would require a strongly anharmonic Hamiltonian
for the nuclear collective modes, which would not be supported
by traditional nuclear models~\cite{BZ93}. Arguments supporting
such anharmonicities have recently been presented in
ref.~\cite{volpe}. Another effect arising from anharmonicity
would be the spin or isospin splitting of the DGDR. Since the
Coulomb interaction favors lower energy excitations, it is clear
that a decrease of the DGDR centroid  would increase its cross
section. A similar effect would occur if a strongly populated
substate is splitted to lower energies.  To study this point, we
have varyied the energy of the DGDR centroid in the range $20 \
{\rm MeV}\le E_{DGDR} \le 27 \ {\rm MeV}$. The obtained DGDR
cross sections (including direct excitations) are equal to 620
mb, 299 mb and 199 mb, for the centroid energies of 20 MeV, 24
MeV, and 27 MeV, respectively. Although systematics of the
DGDR excitation \cite{hans,chomaz} do not show large
deviation of the centroid energy, the data are not
conclusive, and more experiments are clearly necessary. We
conclude, that from the arguments analysed here, the
magnitude of the DGDR cross section is more sensitive to the
energy position of this state. The magnitude of the DGDR cross
section would increase by a factor 2 if the energy position of
the DGDR decreases by 20\%, as found in ref. \cite{volpe}.

\bigskip

\section{Discussion and conclusions}

In this paper we investigated at great length the excitation of
giant resonances in heavy-ion reactions. Both the single- and
double- giant dipole resonances were considered. The effect of the
finite lifetimes of these resonances on their excitation probabilities
was carefully assessed. The comparison with the available experimental
data shows that some physics is still missing. Here we address this
issue.

In our discussion of the excitation of a damped giant resonance,
the damping arises from the coupling to the large number of
non-collective states that surround the GDR and shares with it
its quantum numbers. One should keep in mind that our
final result for the excitation probability involves an implicit
average over the ``chaotic" degrees of freedom whose quantum
manifestition is just the fine structure states. At this point
one is reminded of a well known fact in reaction theory, namely, ensemble
or energy averaged cross sections contain two pieces: one obtained
from an average amplitude, or ``optical" piece, and a second piece which
arises from the fluctuations. We expect similar contribution of the
fluctuations to the excitation probability in the case of the GR. Here,
however, the fluctuations are in the ``host" nucleus and not in the
compound nucleus.

At this point, we recall similar type of fluctuations
which constitute the dominant piece in the case of deep inelastic
heavy ion reactions~\cite{Brink}, when it is assumed that only chaotic
channels are involved in the inelastic transitions. The investigation
of the effects of fluctuations on the excitation of giant resonances
in heavy-ion reactions, following the procedure of ~\cite{Brink}, is
underway and will be reported in a future publication.

\bigskip\bigskip

{\bf Acknowledgements}

We would like to express our
gratitude to Dr. Hans Emling for useful comments and suggestions during
the development of this work. One of us (CAB) acknowledges finantial
support from  the GSI-Darmstadt. This work was supported in part by
the Brazilian agencies: CNPq and FINEP. This work was also partially
supported by the National Science Foundation through a grant for the
Institute for Theoretical Atomic and Molecular Physics at Harvard
University and Smithsonian Astrophysical Observatory.

\newpage
\noindent
{\bf Figure Captions}  \\
\begin{description}
\item[Fig. 1 - ] Schematic representation of the excitation of Giant Resonaces,
	       populated in heavy ion collisions.
\item[Fig. 2 - ] Time-dependence of the occupation probabilities $|a_0|^2$ and
	       $|a_1|^2$, in a collision with impact parameter $b=15$~fm.
	       The time is measured in terms of the dimensionless variable
	       $\tau=(v\gamma/b)\ t$.
\item[Fig. 3 - ] The GDR excitation probabilities as functions of the impact
	       parameter, for sharp and smooth absorptions.
\item[Fig. 4 - ] Nuclear excitation probabilities as  functions of the
	       impact parameter.
\item[Fig. 5 - ] Excitation energy spectra  of the main Giant Resonances for
	       both Breit-Wigner and Lorentzian line shapes.
\item[Fig. 6 - ] Ratio between the DGDR and the GDR cross sections as a
function
	       of the bombarding energy.
\item[Fig. 7 - ] Dependence of $\sigma_{\sc GDR}$ and $\sigma_{\sc DGDR}$ on
the
	       GDR width, treated as a free parameter. For details
	       see  the text.
\end{description}
\newpage

Table 1.
Parameters \cite{Ra79} for the nucleon-nucleon amplitude,
$f_{NN}(\theta=0^\circ)=
(k_{NN}/4\pi)$ $\sigma_{NN} \ (i+\alpha_{NN})$.

\begin{center}
\begin{tabular}{|l|l|l|l|l|l|l|r|} \hline\hline
E [MeV/nucl]&$\sigma_{NN}$ [fm$^2$]&$\alpha_{NN}$ \\ \hline
85&6.1&1 \\ \hline
94&5.5&1.07 \\ \hline
120&4.5&0.7 \\ \hline
200&3.2&0.6 \\ \hline
342.5&2.84&0.26 \\ \hline
425&3.2&0.36 \\ \hline
550&3.62&0.04 \\ \hline
650&4.0&-0.095 \\ \hline
800&4.26&-0.075 \\ \hline
1000&4.32&-0.275 \\ \hline
2200&4.33&-0.33 \\ \hline\hline
\end{tabular}
\end{center}
\bigskip
\bigskip

Table 2 : Excitation cross sections (in milibarns) of the IVGDR,
and of the $n\times GDR$ states in the reaction $^{208}Pb+^{208}Pb$
at 640 MeV.A. A comparison with first order perturbation theory and the
harmonic oscillator is made.

\begin{center}
\begin{tabular}{|l|l|l|r|} \hline\hline
State&1st pert. th.&harm. osc.&c.c. \\ \hline\hline
$IVGD$&3891&3235&3210 \\ \hline
$2\otimes IVGD$&388&281&280 \\ \hline
$3\otimes IVGD$&39.2&27.3&32.7 \\ \hline
$4\otimes IVGD$&4.2&2.4&3.2 \\ \hline\hline
\end{tabular}
\end{center}

\bigskip\bigskip

Table 3 :  Transition probabilities at $b=14.3\ fm$, for the
reaction $^{208}Pb+^{208}Pb$ at 640 MeV.A.
A comparison with first order perturbation theory is made.

\begin{center}
\begin{tabular}{|l|l|r|} \hline\hline
Trans.&1st pert. th.&c.c. \\ \hline\hline
$g.s.\longrightarrow g.s.$&---&0.515 \\ \hline
$g.s.\longrightarrow IVGD$&0.506&0.279 \\ \hline
$g.s.\longrightarrow ISGQ$&0.080&0.064 \\ \hline
$g.s.\longrightarrow IVGQ$&0.064&0.049 \\ \hline
$g.s.\longrightarrow 2\otimes IVGD$&0.128&0.092 \\ \hline\hline
\end{tabular}
\end{center}
\newpage

\bigskip\bigskip

Table 4 :  Centroid energies and widths of the main Giant Resonances in
	   $^{208}$Pb.
\begin{center}
\begin{tabular}{|l|l|l|l|l|r|} \hline\hline
&GDR&DGDR&ISGQR&IVGQR \\ \hline\hline
$E_r$ (MeV)&13.5&27.0&10.9&20.2 \\ \hline
$\Gamma$ (MeV)&4.0&5.7&4.8&5.5 \\ \hline\hline
\end{tabular}
\end{center}

\bigskip\bigskip

Table 5 :  Cross sections in milibarns for the excitation of giant resonances
in lead, for the
reaction $^{208}Pb+^{208}Pb$ at 640 MeV.A.
\begin{center}
\begin{tabular}{|l|l|l|l|r|} \hline\hline
GDR&DGDR&ISGQR&IVGQR \\ \hline\hline
2704&184 (199) [198]&347&186 \\ \hline\hline
\end{tabular}
\end{center}
\end{document}

================== RFC 822 Headers ==================
Received: from ntsu1.if.ufrj.br by nthp2.if.ufrj.br with SMTP
	(1.38.193.4/16.2) id AA17737; Thu, 13 Jul 1995 20:22:28 -0300
{}From: bertu@if.ufrj.br
Received: (bertu@localhost) by ntsu1.if.ufrj.br (8.6.9/8.6.9) id UAA14996 for
bertulani@vscn.gsi.de; Thu, 13 Jul 1995 20:27:03 -0300
Date: Thu, 13 Jul 1995 20:27:03 -0300
Message-Id: <199507132327.UAA14996@ntsu1.if.ufrj.br>
To: bertulani@vscn.gsi.de
Subject: x